# Monetary Policy and the Gendered Labor Market Dynamics: Evidence from Developing Economies

**Marjan Petreski**
University American College Skopje
marjan.petreski@uacs.edu.mk

**Stefan Tanevski**
University American College Skopje
stefan.tanevski@uacs.edu.mk

**Alejandro D. Jacobo**
Universidad Nacional de Córdoba
alejandro.jacobo@unc.edu.ar

## Abstract

Using a Taylor rule amended with official reserves movements, we derive country-specific monetary shocks and employ a local projections' estimator for tracking gender-disaggregated labor-market responses in 99 developing economies from 2009 to 2021. Results show that women experience more negative post-shock employment responses than men, contributing to a deepening of the gender gaps on the labor market. After the shock, women leave the labor market more so than men, which results in an apparently intact or even improved unemployment outcome for women. We find limited evidence of sector-specific reaction to interest rates. Additionally, we identify an intense worsening of women's position on the labor market in high-growth environments as well under monetary policy tightening. Developing Asia and Latin America experience the most significant detrimental effects on women's employment, Africa exhibits a slower manifestation of the monetary shocks' impact and developing Europe shows the mildest effects.

1. Introduction

There is a growing interest in analyzing the various consequences of monetary policy driven by the need to not only better understand its impact on total employment but to also identify —and comprehend— the labor market channels underlying its distributional effects. As to this second and novel driver, the rising levels of income inequality in recent decades have made distributional issues a key concern for the general public as well as for economic policymakers. These include central bankers, who have argued if —and how— monetary policy affects the distribution of incomes and whether these distributional effects should be considered. In words of Mersch (2014), "all economic policy-makers have some distributional impact as a result of the measures they introduce —yet until relatively recently, such consequences have been largely ignored in the theory and practice of monetary policy". Draghi (2016) has also exhibited his worries about the distributional effects of monetary policy when he discussed in his past remarks why interest rates were so low at that time, and what the implications of those low rates really were.

Undoubtedly, the distributional effects of monetary policy are complex and uncertain (Bernanke, 2015). Determining these effects is intricated because monetary policy affects individuals' incomes through a large number of channels, many of which are likely to have opposite effects for the distribution of their incomes. However, to properly understand the distributional effect of monetary policy well worth the effort. Otherwise, policymakers will sincerely look as "innocent bystanders" along the different channels via which the monetary policy shocks affect inequality (Coibion et al., 2017).

Despite the prolonged prevalence of dual mandates as primary targets for central banks in advanced economies for over two decades, the impact of monetary policy shocks onto labor market gender gaps has recently captured academic and central bankers attention (Flamini et al. 2023). This interest is grounded on a dual imperative: firstly, to comprehensively grasp how monetary policy affects overall employment as well as economic output and secondly, to pinpoint specific channels within the labor market that contribute to the distributional effects of monetary policy.

An increasing number of studies have pushed the Sisyphean boulder and focused on how monetary policy affects different sectors of the labor market (see for example Singh et al., 2022), occupational groups or labor income (Gomes et al., 2023; Madeira and Salazar, 2023; Amberg et al., 2022; Dolado et al.; 2021; Heathcote et al., 2020; or Zens et al., 2020), and the labor market gender gap. As to the latter, a monetary shock —often thought of as gender-neutral one— tends to influence women and men differently because of their diverse paid and non-paid positions in the economy.

In fact, on the one hand, men and women are not equally represented across sectors and jobs. Men are more likely than women to work in construction and manufacturing industries which according to Erceg and Levin (2006) are more sensitive to changes in interest rates than non-durable services, where more women are usually employed. Even within services, women tend to work in areas such as education and healthcare that are less sensitive to economic fluctuations following monetary policy changes. On the other hand, women are also more likely to be employed in jobs that that are more disposed to labor market adjustments due to monetary policy changes (part-time or temporary contracts come here as examples). Entrepreneurial women also tend to be primary caregivers and are more likely to reduce their labor force participation in turbulent times (Takhtamanova and Sierminska, 2009). Due to vulnerable positions in the labor market, monetary policy may definitively not be gender neutral. Still, scarce attention to these issues has been given by monetary policies and, as a consequence, these policies have seldom been conducive to the achievement of gender equality.

The aim of this study is to investigate if heterogeneous monetary policy shocks have a tendency to be 'gender biased'. From this viewpoint, we depart from the focus on the policy shocks' effects over income and/or occupational groups in that we center on another unexplored aspect - gender gaps on the labor market, particularly analyzing how monetary policy shocks impact men's versus women's employment, in which sectors and through which adjusting process.

To our knowledge, the relationship between monetary policy shocks and gender employment gaps has rarely or never been examined as we do. Flamini et al. (2023) only studies the relationship for a set of OECD countries. As a key contribution, we derive country-specific monetary shocks and employ a local projections' estimator for tracking gender-disaggregated labor-market responses in 99 developing economies from 2009 to 2021. As another novelty, we further advance in the methodological approach by considering official reserves' changes in the monetary policy function, as a way to capture the characteristic of developing economies which more frequently run forms of rigid exchange rates and/or heavily intervene on the foreign exchange market to prevent large volatilities in prices and outputs. Hence, if the conduct of monetary policy through sterilized forex interventions is not captured in the model, it would be improperly identified as a monetary policy shock, while potentially being a daily monetary-policy management in countries with fixed exchange rates or currency boards. We consider this is the first paper to conduct such an analysis for a set of developing economies.

The rest of the paper is organized as follows. Section 2 provides a brief overview of the referent literature. Section 3 presents the underlying methodology and the data used, portraying all the constraints one usually faces when working with developing economies. Section 4 presents the results. The last section concludes.

## 2. Overview of the related literature

Depending on different features, gender as well as racial minorities, are extremely affected by contractionary monetary policy. While monetary policy affects gender labor market gap, the sign of the effect continuous to be unclear, however. The scant empirical evidence suggesting that women's labor market outcomes may be more vulnerable to monetary policy shocks than men's has also not been conclusive since Takhtamanova and Sierminska (2009) find no significant impact of monetary policy changes on gender gaps in employment for OECD countries.

Examining the dissimilar impacts of monetary policy on unemployment rates in the United States, Abell (1991) concluded that the labor market is ghettoized —the term is ours— in a way that tends to favor white men during periods of contractionary monetary policies. Likewise, Thorbecke (2001) found similar results indicating that disinflationary monetary policy increases unemployment among minorities approximately twice as much as it does among whites. Carpenter and Rodgers (2004) highlighted that monetary policy appears to have a disproportionate effect on the unemployment rate of teenagers for example, particularly African American ones, and show that a monetary policy accommodation reduces the gap between the unemployment rates of black and white households.

Braunstein and Heintz (2008) also consider the employment costs of inflation reduction in developing countries from a gender perspective and explore two broad empirical questions: (1) what is the impact of inflation reduction on employment, and is the impact different for women and men, and (2) how are monetary policy indicators (e.g. real interest rates) connected to deflationary episodes and gender-specific employment effects? Their study reveals that the gap between women's and men's employment increases when central banks tighten monetary policy to lower inflation in emerging markets and developing countries. Similarly, but for the United States, Seguino and Heintz (2012)' results indicate that the costs of fighting inflation are unequally distributed amongst workers. For these authors, the effects vary according to the density of the black population in each US state and that the cost of policies to combat inflation is unevenly distributed among workers, negatively affecting more heavily on Black women and Black men, followed by white women and lastly white men.

Differences in unemployment rates across groups seem to be, in fact, most pronounced during an economic downturn and disappear throughout an expansion. A sustained expansion excessively improves labor market outcomes for the most susceptible groups of workers in the United States (Duzak, 2021), for whom labor market sensitivities also vary across gender and racial groups. In this country, Black and Hispanic workers face the most adverse impact from economic slowdowns, especially men. Recall that gender and racial discrimination may be complements, such that white women, Black women, and Black men all face relatively similar disadvantages in job access during economic downturns (Seguino and Heintz, 2012).

In an interesting document, Bartscher et al. (2022) link monetary policy shocks not only to earnings but also to wealth differentials between black and white households. They find that while accommodative monetary policy tends to reduce racial unemployment and thus earnings differentials —and, by the way, it exacerbates racial wealth differentials— which implies an important tradeoff for policymakers.

Bergman et al. (2022) find that women tend to increase their employment more than men under expansionary monetary policy in tighter labor markets. They show that the employment of populations with lower labor force attachment (blacks, women and high school dropouts) is more responsive to expansionary monetary policy in tighter labor markets.

Among those who explore the monetary policy impact on gender gap, the majority of the works focus on advanced economies. As to developing ones, where variables as gender and race become also indispensable in the debate for a more strategic economic policy, the studies are occasional. Beyond the study of Braunstein and Heintz (2008) who analyze 17 low- and middle-income countries, Couto and Brenck (2024) explores the effect of changes in the interest rate for female and black employment creation in Brazil. The authors concludes that social stratification, if not considered, can lead to misleading policies that perpetuate unequal socioeconomic outcomes. This is because the real interest rate has a positive effect on the relative unemployment of Black men to white men, no effect on the relative unemployment of Black women to white men, and a negative effect on the relative unemployment of white women to white men.

To our knowledge, the latest contribution trying to shed some light on how monetary policy affects gender employment gaps in a panel of 22 advanced and emerging market economies is the work of Flamini op. cit. The authors analyze how exogenous monetary policy shocks impact women's employment versus men's in which sectors, through which adjustment process (labor force participation and unemployment rates), and how different labor market characteristics shape these effects. They also study the asymmetric effects of contractionary versus expansionary monetary policy shocks and across business cycles (recessions versus expansions). Their results show that men's employment falls more than women's after contractionary monetary policy shocks, narrowing the employment gender gap over time. The effects are larger in countries with more flexible labor market regulations, higher gender wage gaps, and lower informal women's employment compared to men's. Finally, the effects are also larger for contractionary monetary policy shocks and during expansions.

However, beyond the above-mentioned studies, the gender impact of unanticipated monetary policy shocks on labor markets in developing economies remains unexplored. Probably this is due to several reasons. It is not the scope of this paper to specify a complete explanation of this circumstance and we are not going into further details regarding this lack of studies. However, it may perhaps be noted without straying too far afield from our major focus that this exploration requires a proper definition of the monetary shock, which tends to be a difficult task in developing economies where the financial instability has also been an important characteristic.

As to this point, it is true that most of these economies has been modernizing their monetary policy frameworks, often moving toward an inflation targeting monetary policy. However, questions regarding the strength of monetary policy transmission from interest rates to inflation and output have been delayed. The growing concerns in recent years about financial stability raise the question whether central banks could pursue such a goal, and if so, how. Not surprisingly, a large body of literature on central banks actions focuses on the inclusion of various kinds of stability measures in the Taylor rule. Formally speaking, one can augment equation with a term related to some measure of financial stability with the accurate weight. However, what exactly this extension of the rule should look like remains an open question. Or, in another words, although there are now numerous papers that present augmented Taylor rules, it is unclear which of those measures would be best to safeguard financial stability (Käfer, 2014), in particular in those economies with rather inflexible forms of exchange rates as the developing ones.

In fact, since the exchange rate determines the price of imported goods as well as inflation expectations and the competitiveness of domestic firms is persuaded by the exchange rate, an appreciation in the domestic currency makes foreign products cheaper and domestic products more expensive. Accordingly, the demand for domestic products should fall in this case. But these two impacts are only linked to the traditional arguments of the Taylor rule: inflation and output. By the way, in this connection, the literature shows that the sectoral composition of labor is also an important channel in which the exchange rate affects gender and race inequality. Indeed, a devaluation would boost exports and it may also affect inflation (Ha, Stocker, and Yilmazkuday, 2020, among many others), and Erten and Metzger (2019) also highlight the importance of the country's sectoral composition and stages of development where a currency undervaluation can have different effects, reducing women's labor force participation by allocating resources to male-dominated, technologically intensive industries.

As to financial instability, capital flows induced by the exchange rate can generate credit and asset price bubbles, and a collapse in the inflowing country. Besides, if the debt weight of firms and banks are to a large extent denominated in a foreign currency, an exchange rate depreciation may increase the burden of outstanding debt and eventually force the economy to a crash. Thus, one may conclude that these economies are most affected by such anxieties as they are usually heavily dependent on exchange rate movements (Ho and McCauley, 2003; Mohanty and Klau, 2005; Aizeman et al., 2011). Not surprisingly, the normative literature mostly suggests small reactions of the interest rate to the exchange rate. This finding seems to be supported by the positive literature, as this usually states significant, albeit rather. small responses (Käffer, op. cit.). While as a first suggestion an exchange rate objective for the ECB would be inappropriate as the Eurozone as a whole is anything, it seems reasonable in a small and emerging and/or dollarized de facto economy. Then, the Taylor rule reactions to the exchange rate proceeds (Ball, 1999; Svensson, 2000; Batini et al., 2003), it should be amended and the way in we do so will be unveiled in the next section.

## 3. Methodology and data

Our approach for identifying monetary policy shocks onto labor market outcomes for a set of developing economies consists of two parts. In the first one, we identify monetary policy shocks by estimating an adjusted Taylor rule, following Brandao-Marques et al. (2020). We index the countries by $k$ and years by $t$. Let $i_{k,t}$ represent the short-term central-bank nominal interest rate, $g_{k,t}$ the GDP growth rate, $\pi_{k,t}$ denote the inflation rate and $f_{k,t}$ the change in central-bank foreign exchange reserves. The superscript $F$ denotes one year ahead forecast for the GDP growth and inflation. The rest of the variables are taken with their first lags, inter alia to suppress any endogeneity concerns. We employ an OLS in estimating the Taylor-type regression for each country separately:

$$i_{k,t} - i_{k,t-1} = \alpha_{0,k} + \alpha_{1,k} g^F_{k,t+1} + \alpha_{2,k} \pi^F_{k,t+1} + \alpha_{3,k} g_{k,t-1} + \alpha_{4,k} \pi_{k,t-1} + \alpha_{5,k} f_{k,t-1} + \alpha_{6,k} i_{k,t-1} + \varepsilon_{k,t} \qquad (1)$$

Differently than the original Taylor rule and as a novelty, we add the change in central bank reserves to account for the fact that, most of the developing economies are small and open economies who either follow rigid forms of their exchange rate regimes or regularly intervene on the foreign exchange markets to prevent large swings in the exchange rate be translated into large volatility of prices and output. Then, this may be a critical constraint or determinant in the monetary policy conduct. Such developments would not be captured by taking the nominal exchange rate and is clearly recognized in Brandao-Marques et al. (2020). By so doing, we believe we ameliorate this problem. Still, the residual may still capture exogenous variation (purged from any impact of the lagged values in the included variables).

All coefficients are country specific at this stage, while panel estimates of equation (1) are provided in **Table A 1** in Appendix 1 for intellectual curiosity. Monetary policy shocks in (1) are identified as the estimated residuals $\varepsilon_{k,t}$, i.e. through the deviations from the Taylor rule which aim to capture the unanticipated and non-systematic components of monetary policy actions. As the magnitude of shocks varies significantly across countries, we standardize the residuals on a country-by-country basis. Consequently, a unit monetary policy shock represents a one standard deviation shock within each specific country.

Inflation rates, GDP growth rates as well as their forecasts are obtained from the International Monetary Fund's (IMF) World Economic Outlook (WEO). Forecasts refer to next-year forecasts published each October, hence for example the 2021 inflation forecast is the one published in the WEO October 2020 edition. This implies that if IMF published any revision after this date, this is not taken into account. The short-term interest rate is sourced from the International Financial Statistics (IFS), but to serve as large sample as possible, the lending interest rate is taken, given the sample with the policy interest rate has been considerably smaller. However, this has the advantage in that the lending rate may better capture monetary policy stance in small and open economies with more rigid forms of exchange rates, if they used other instruments of the monetary policy, like the reserve requirement or forex operations to manage domestic liquidity; or in some cases, countries who are dollarized/euroized and where a policy rate is not even existent. This further addresses any remaining exogenous variation in the residuals of (1), stemming from the usage of other monetary policy instruments, which is to our advantage, however it may contain exogenous variation which is bank-sector specific, which is to our disadvantage. The reserves are taken from the IFS in their current dollar value. All variables' definitions and sources are provided in **Table A 2**, while descriptive statistics in **Table A 3** in the Appendix 2.

For this first part of the analysis, we commence by taking all countries which the IMF classifies as emerging market and developing economies, a total of 160. Our selection is the entire period after the Global Financial Crisis of 2007-08, i.e. starting in 2009 and ending in 2022. However, the maximal number of 2,240 observations is trimmed to 975 determined by missing observations (being usually the case for the small island countries or the countries in conflict), usage of lags in our own specification, as well as by dropping the countries for which fewer than five years within the specified span were available. Therefore, we are left over with a total of 99 developing economies. These are specified in **Table A 4** in the Appendix 2.

In the second part of our analysis, we estimate the responses of specific labor market outcomes to monetary policy shocks, following Jorda (2005)'s and Flamini et al. (2023)'s local projections approach for which we use the already estimated policy shock series (lagged) from equation (1), $\varepsilon_{k,t-1}$. In our empirical model, we disentangle the outcome variable, $y_{n,k,t+h}$, by gender, and then take is a gender gap. The following are used as outcome variables: employment rate of working-age population (15+) and of youth (15-24), share of employment in agriculture, industry and services, labor force participation rate and unemployment rate. The gender gap is quantified by subtracting the value for men from the value for women for each variable. Additionally, we introduce the notation 'h' to represent the horizon of the estimated responses, spanning up to five years (h = 0, ..., 5), following the shock at time t-1. Let $\lambda_{k,h}^n$ denote country fixed effects, and $\theta_{t,h}^n$ represent time fixed effects. For each horizon h, a distinct fixed-effects panel regression is estimated as follows:

$$y_{n,k,t+h} = \beta_{n,h}\varepsilon_{k,t-1} + \alpha_{n,h}y_{n,k,t-1} + \lambda_{k,h}^n + \theta_{t,h}^n + v_{n,h,k,t} \qquad (2)$$

The estimated coefficient $\beta_{n,h}$ provides a measure of the percentage (point) change at horizon h, reflecting the response to a monetary policy shock of one standard deviation. To visually depict these findings, we construct graphical representations by plotting the estimated coefficients along with their confidence intervals on the vertical axis, aligning them against their corresponding horizons on the horizontal axis.

We conduct a few subsequent steps to observe heterogenous results and/or to provide some robustness analysis. First, to test whether the impact of the monetary policy shocks depends on the economic conditions as defined through the real GDP growth, to the explanatory variables in (2), we add the lag of the real GDP growth as follows:

$$y_{n,k,t+h} = \beta_{n,h}\varepsilon_{k,t-1} + \alpha_{n,h}y_{n,k,t-1} + \rho_{n,h}g_{n,k,t-1} + \lambda_{k,h}^n + \theta_{t,h}^n + v_{n,h,k,t} \qquad (3)$$

Then, we take an alternative specification of our Taylor rule (equation 3). Namely, we calculate the residuals of short-term interest rate forecast errors after controlling for GDP and CPI forecast errors instead of their forecasts.

Second, in equation (3), we add labor conditions, represented through three variables: collective bargaining coverage rate, gender pay gap and the informal employment share, introduced through the vector $X'_{n,k,j,t-1}$:

$$y_{n,k,t+h} = \beta_{n,h}\varepsilon_{k,t-1} + \alpha_{n,h}y_{n,k,t-1} + \rho_{n,h}g_{n,k,t-1} + \sum \gamma_{n,h,j}X'_{n,k,j,t-1} + \lambda_{k,h}^n + \theta_{t,h}^n + v_{n,h,k,t}$$
$$(4)$$

in order to observe if some labor-market adjustment could help in explaining differential effects for men and women of a monetary policy shock.

Third, to test asymmetric impacts of monetary policy shocks onto gender gaps in the labor market, we run the following adjusted model:

$$y_{n,k,t+h} = \beta_{n,h}^-\varepsilon_{k,t-1}G(d_i) + \beta_{n,h}^+\varepsilon_{k,t-1}(1 - G(d_i)) + \alpha_{n,h}y_{n,k,t-1} + \rho_{n,h}g_{n,k,t-1} + \lambda_{k,h}^n + \theta_{t,h}^n + v_{n,h,k,t} \qquad (5)$$

whereby $G(d_i) = \frac{\exp(-\eta z_i)}{1+\exp(-\eta z_i)}, \eta > 0$. $z_i$ is a normalized indicator of the mean state of the country to capture cross-country variation defined as $z_i = \frac{x_i - \bar{x}}{\sigma_x}$, whereby $x_i$ stands for the country average while $\bar{x}$ and $\sigma_x$ the cross-country average and standard deviation, respectively. We use $\eta = 1.5$ (following Auerbach and Gorodnichenko, 2013) to estimate coefficients $\beta$s as the percentage (point) changes at horizon $h$ in response to a monetary policy shock of one standard deviation in low versus high real GDP growth regimes. In the second, $G(d_i)$ is reduced to a dummy variable that takes a value of one for positive monetary policy shocks and zero otherwise, to quantify the gendered labor-market outcomes during positive and negative monetary policy shocks.

Finally, we run equation (3) for different geographical subset of countries as follows: Emerging and Developing Asia (EDA), Emerging and Developing Europe (EDE), Latin America and the Caribbean (LAC), Sub-Saharan Africa (SSA). The belonging of each country is given in **Table A 4** in Appendix 2. Note that the Middle East countries are included under Emerging and Developing Asia, due to the fairly small sample to obtain results separately.

Labor market data used for our outcome variable, as well as the three variables capturing the labor conditions: collective bargaining rate, gender pay gap and the informal employment share, are sourced from the International Labor Organization (ILO). The dataset on the outcome variables is generally richer in terms of country and period coverage as compared to the dataset in our first part of the analysis, i.e. data for 157 countries. However, given our monetary policy shocks are identified for a smaller set of countries/periods, we are confined to the 924 observations introduced above. On the other hand, data on collective bargaining rate, gender pay gap and the informal employment share are significantly scarcer and are available only for about a third of our final dataset. Hence, equation (5) does not incorporate them, despite run after introducing them in the story.

## 4. Results and discussion

This section presents the results and offers a discussion in the following order: baseline results, robustness check with alternative specification of monetary policy shocks, results with using labor-market adjustment variables, results with asymmetries, and geographically-differentiated results.

### 4.1. Baseline results

A monetary policy shock of one standard deviation works differently for men and women in the developing economies (**Figure 1**). Note that the figure is structured so that each row has one labor-market outcome variable, while each column represents, respectively, men, women and the gender gap (calculated as 'women' minus 'men'). Interestingly to note, the response of men's employment outcome to a monetary policy shock is slightly positive in all the employment variables used, though clearly not very different than zero. On the other hand, women's employment reacts to a monetary policy shock in a mixed manner, despite a frequent and signiticativelly negative reaction.

A monetary policy shocks of one standard deviation results in a decline in female employment (15+) of the magnitude of about 0.07 percentage points in the third and fourth year after the shock (row 1). Still, if men's employment reaction could be interpreted as significant and positive, it may indicate that a monetary policy shock prompts men – who are usually the main breadwinner in most of the developing-country societies – to more actively seek employment given that women are more frequently losing their jobs (and incomes) in such circumstances. This determines that the gender employment gap declines over the entire horizon, but starts negligibly and deepens around the third and fourth year at near 0.1 percentage point, after which the shock's effect onto the gap vanishes. This prime result is of similar absolute magnitude, yet smaller and with the opposite sign than that of Flamini op. cit. who find a response of +0.5 percentage points (despite they do not use standardized monetary policy shock, which may impose important differences).

Similar general conclusions could be drawn by observing the rest of the employment variables. For example, the pattern of reaction in the case of youth (15-24) is similar (row 2), yet the negative effect of the monetary policy shocks comes sooner, i.e. already after a year of the shock, imposing a worsening of the gender employment gap. However, women's sectoral employment shares remain somehow intact (rows 3-5), suggesting that a monetary policy shock does not impose sectoral relocation of the female employment, i.e. that the reduction in employment for women tends to be sector-neutral. However, men in agriculture increase their share mainly at the expense of those in services, which suggests that when a monetary policy shock hits, jobs for men are lost in services (this may be associated with service branches as transport and hospitality), but then they find shield in agriculture. On the other hand, women's share remains intact dominating in service branches as trade and public administration, as well in agriculture whereby they more frequently appear as unpaid family members. This implies that women worsen their relative presence in agriculture and improve it in services.

Labor force participation rate reacts in a similar fashion as employment (row 6). Men increase their participation rate following a monetary policy shock, likely reflecting the notion that new employment is also driven by labor-market activation. Women clearly passivize, with the strongest effect arriving at about the fourth year following the shock, which then results in a worsening of the gender participation gap of about 0.1 percentage points at its peak in years three and four. On the other hand, unemployment rate of men does not react until years four and five (row 7), when it starts declining, consistent with their activation and employment attitude following a monetary policy shock. While, that of women, declines more persistently throughout the entire period, with a magnitude of about 0.05 percent, implying a reduction of the gender unemployment gap (favorable for women in the case of this variable). Given we found women to be more frequently than men exiting the labor market (or passivizing), the declining unemployment rate implies that women have higher propensity to passivize after losing their job than men.

Results are almost replicated when the real GDP growth rate is added as independent variable in the equation (**Figure 2**). The idea is that these results serve as robustness check, which they do. In portraying the reaction of the labor-market outcomes of men and women to monetary policy shocks, the overall stance of the economy may matter, which here is captured through the lagged value of the real GDP growth. The results remain stable.

# Figure 1– Gendered labor market response to a monetary policy shock

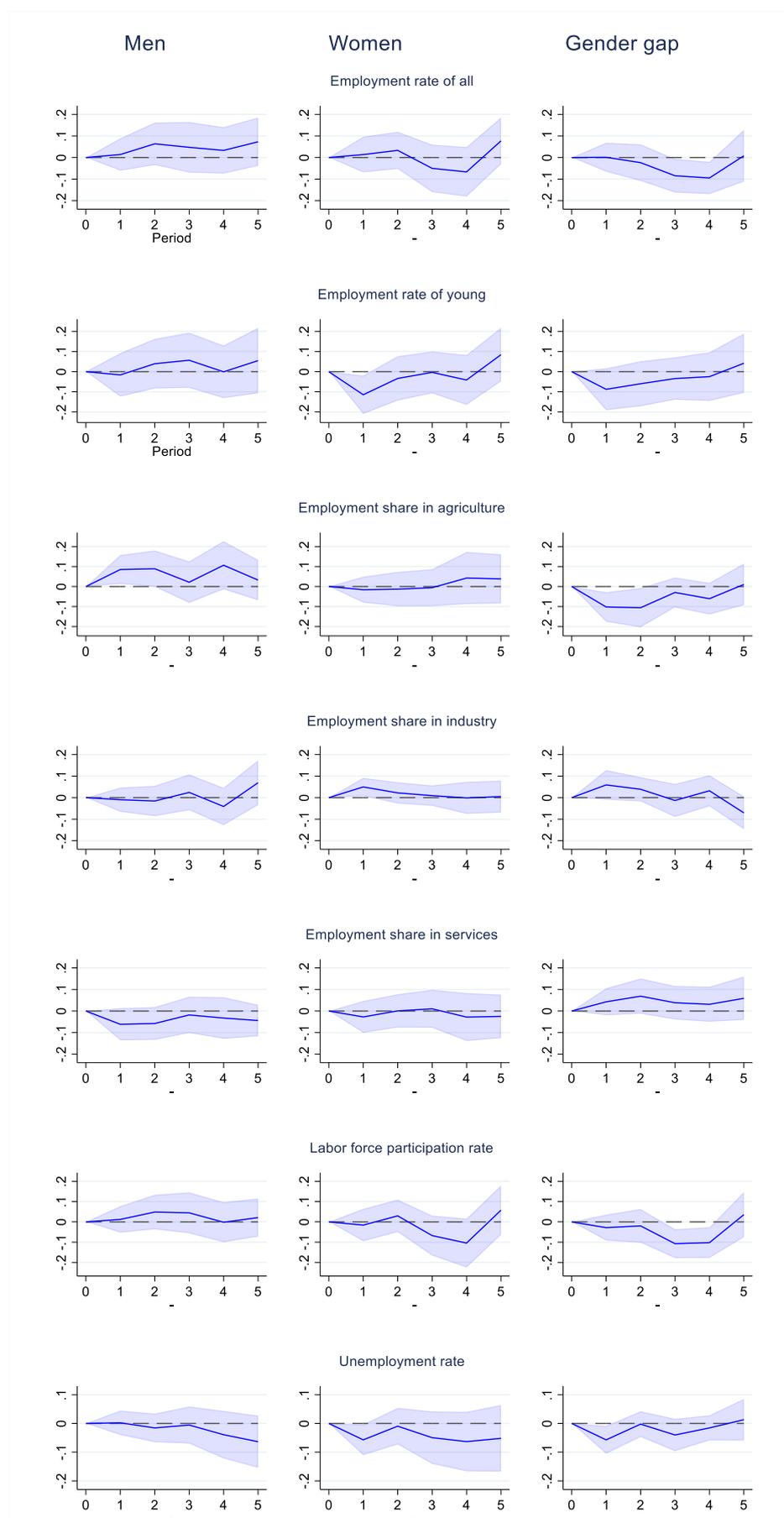

Source: Authors' estimates.

Notes: Each graph presents the response of the titled labor market indicator to a one standard deviation monetary policy shock, separately for men (column 1), women (column 2) and for the gender gap (column 3). Hence, the vertical axis of columns 1 and 2 represent percentages, while of column 3 percentage points. The horizontal axis presents the time horizon expressed in years. 90% confidence interval is presented in shading. A positive (negative) impulse response represents a narrowing (widening) of the gender gap.

**Figure 2 – Gendered labor market response to a monetary policy shock, with lagged GDP as a control**

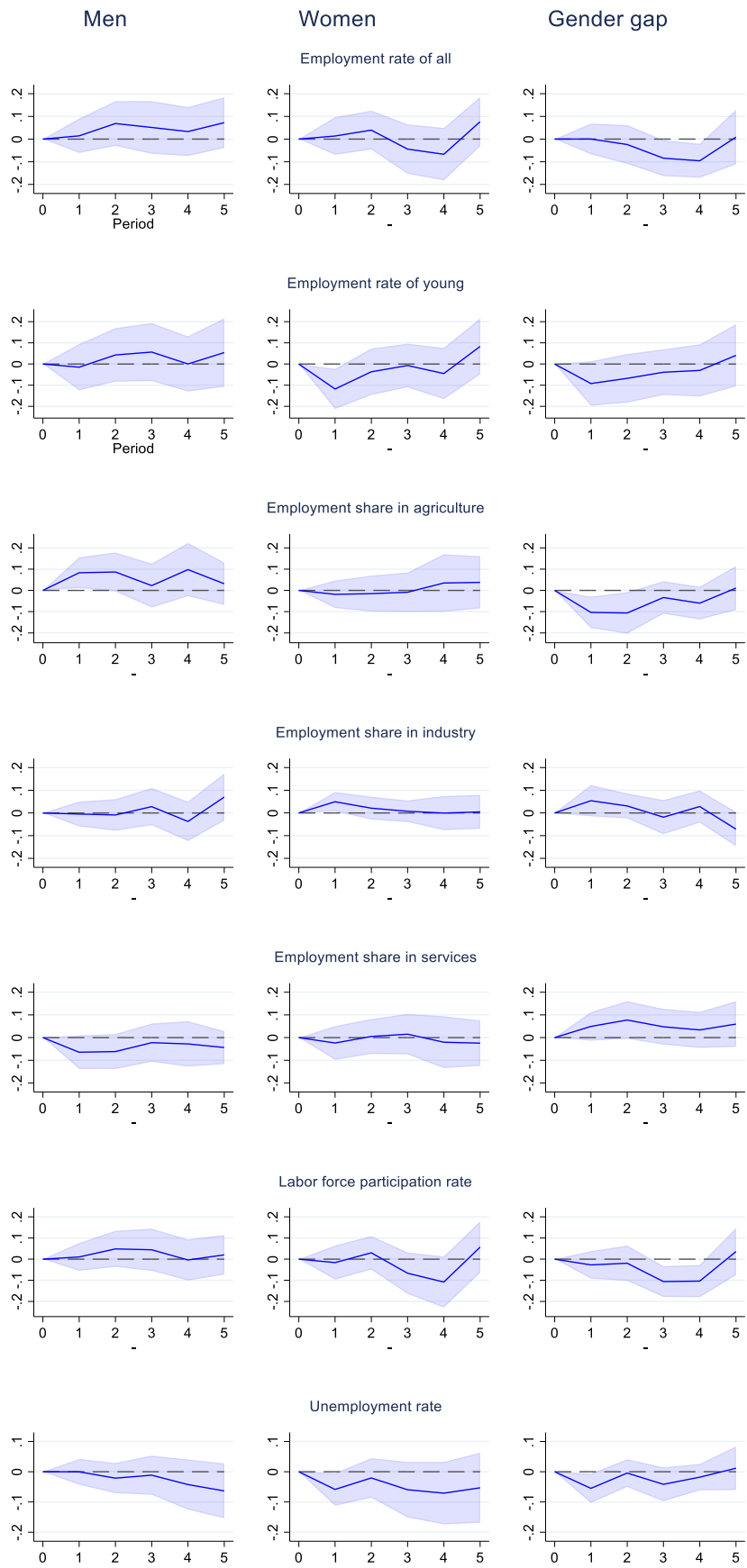

Source: Authors' estimates.

Notes: Each graph presents the response of the titled labor market indicator to a one standard deviation monetary policy shock, separately for men (column 1), women (column 2) and for the gender gap (column 3). Hence, the vertical axis of columns 1 and 2 represent percentages, while of column 3 percentage points. The horizontal axis presents the time horizon expressed in years. 90% confidence interval is presented in shading. A positive (negative) impulse response represents a narrowing (widening) of the gender gap.

### 4.2. Robustness to alternative specification of monetary policy shocks

To test the robustness of our results, we take an alternative specification of our Taylor rule (precisely of equation 5). Namely, we estimate the residuals of short-term interest rate forecast errors after controlling for GDP and CPI forecast errors instead of their forecasts. **Figure 3** presents results comparable to those of our baseline model in **Figure 2**. The response of the men's and women's employment rates is similar, despite slightly more intensive on the positive side, which implies that the gap's reaction is weaker though in the same direction of its worsening. The same holds true for the youth employment variables. Sectoral employment reactions are very similar. In the case of the labor force participation rate, the positive reaction of men's indicator is slightly stronger, while of women is slightly weaker at around year fourth, which implies narrower reaction of the gender gap. In the case of unemployment, women's rate reacts in the same fashion but more intense at the longer horizon, which implies that over the same time period the gap worsens (to the advantage of women in this indicator) rather than remaining intact in the baseline scenario.

**Figure 3 – Gendered labor market response to a monetary policy shock, alternative specification of the monetary policy shocks**

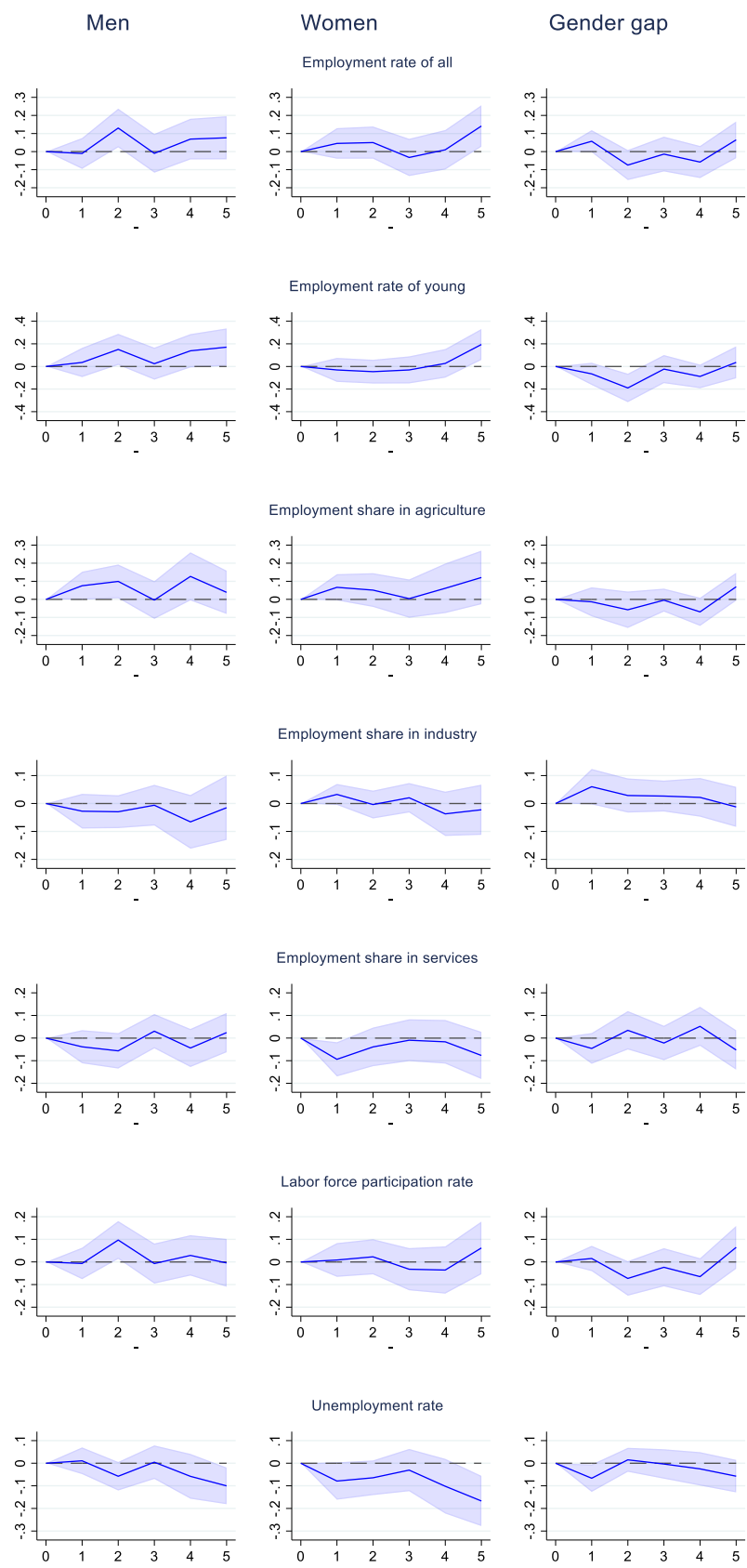

*Source: Authors' estimates.*

Notes: Each graph presents the response of the titled labor market indicator to a one standard deviation monetary policy shock, separately for men (column 1), women (column 2) and for the gender gap (column 3). Hence, the vertical axis of columns 1 and 2 represent percentages, while of column 3 percentage points. The horizontal axis presents the time horizon expressed in years. 90% confidence interval is presented in shading. A positive (negative) impulse response represents a narrowing (widening) of the gender gap.

### 4.3. Results using labor-market adjustment mechanism

The adjustments in the labor market following policy changes are expected to vary based on both the sector's responsiveness to interest rates and the structural characteristics of the labor market. We next control for the labor market adjustment variables in our specification (**Figure 4**). There are more pronounced differences compared to our baseline results, particularly in the case of men's labor market outcomes, which, in addition to capture the labor market adjustment heterogeneity, are driven by the fact that these conditions were not available for about two thirds of our sample. Yet, we consider an advantage that given such severe cut of the sample, results remain fairly robust.

The same pattern of reaction of the gender employment gap (15+) is observed for example, yet significantly stronger: at the deepest point in year four, the gender employment gap worsens by 0.25 percentage points, that is about 2.5 times deeper than in the original specification. However, the confidence interval is likewise wider as expected, probably reflecting the smaller sample.

Then, the reaction of the young gender employment gap changes from negative to slightly positive, reflecting the negative response of the young men employment. Likewise, in the case of the sectoral distribution of employment, the reaction in agriculture is negative, while this reaction is positive in industry and services. This may be related to the dissimilar adjustment mechanisms in the sector; e.g. the fact that collective bargaining coverage rate is significantly smaller in agriculture, or that informal employment there is flagrant, or that the gender pay gap is theoretically infinite due to the prevalence of unpaid family workers among women. This fact, then, attenuates our earlier consideration that women, despite they are less affected sectorally, improve their position in services and worsen it in agriculture. It is safer to say that, given labor sectoral condition, the gender gaps there remain almost intact following a monetary shock.

The reactions of the labor force participation and the unemployment rate are more robust, despite women firstly increase their unemployment rate and then they reduce it, following a monetary policy shock. This is opposed to the baseline scenario when the reaction was mostly on the decline side. As a result, the gender unemployment gap firstly increases (which is detrimental for women), but then its augment is neutralized.

**Figure 4 – Gendered labor market response to a monetary policy shock, with labor conditions as controls**

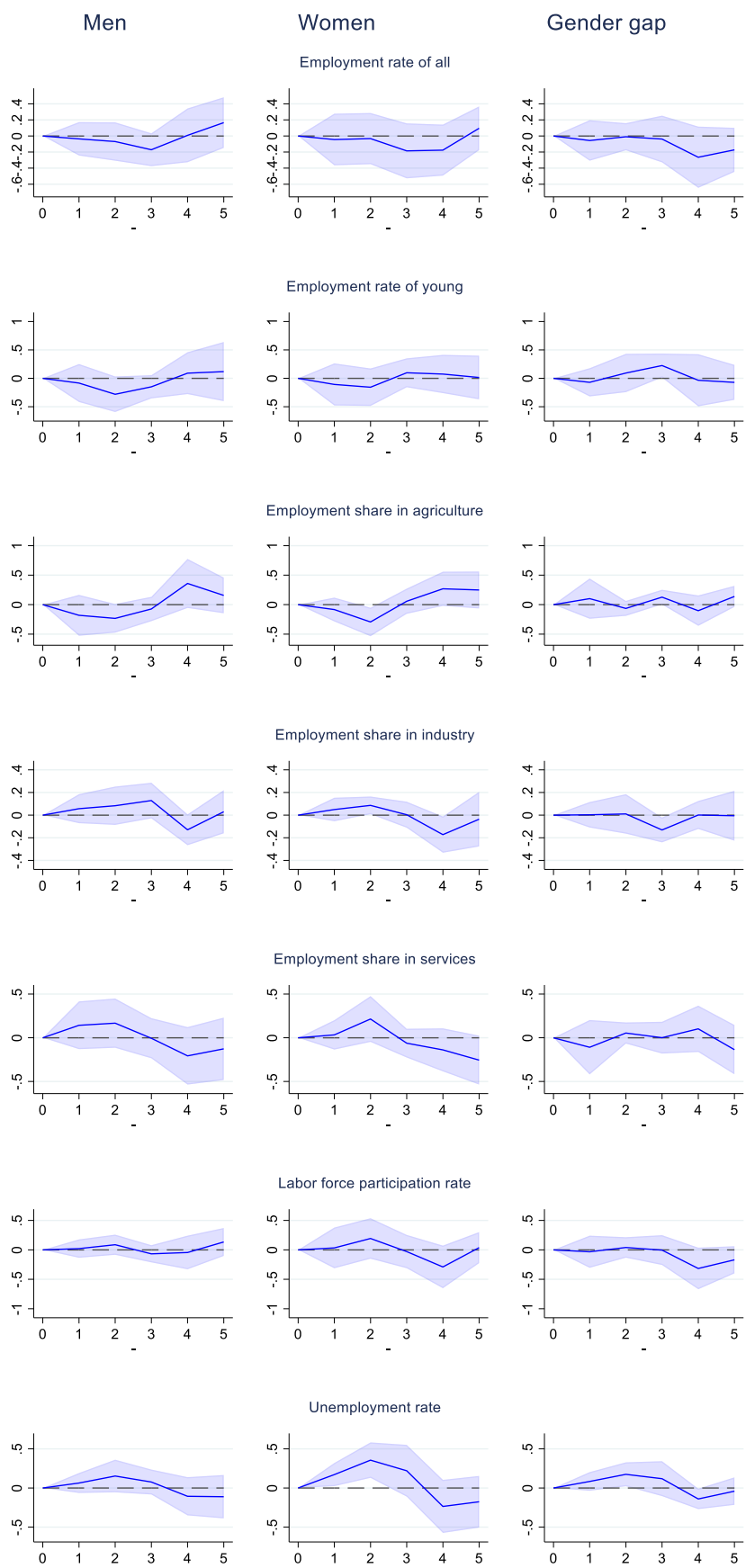

*Source: Authors' estimates.*

*Notes: Each graph presents the response of the titled labor market indicator to a one standard deviation monetary policy shock, separately for men (column 1), women (column 2) and for the gender gap (column 3). Hence, the vertical axis of columns 1 and 2 represent percentages, while of column 3 percentage points. The horizontal axis presents the time horizon expressed in years. 90% confidence interval is presented in shading. A positive (negative) impulse response represents a narrowing (widening) of the gender gap.*

### 4.4. Results with asymmetries

Given the likely importance of the labor market adjustment for the sectoral distribution of employment, in a situation of a severely cut sample due to data constraints related to labor conditions variables, in what follows, we constrain the analysis to the employment rates of 15+, of youth, the labor participation and unemployment rates only.

**Figure 5** divides the countries based on whether their average growth rate of GDP over the observed period has been low or high compared to the cross-country average. The gender employment gap almost does not react to a monetary policy shock in low-growth environment, while it significantly worsens especially in years three and four in a high-growth environment. Similar pattern is observed for the gender employment gap of youth, as well for the labor force participation gap, though worsens following a monetary policy shock in high-growth environment, and then significantly rebounds. On the other hand, the gender unemployment gap mainly reacts positively though at a longer run; this implies that women experience faster-growing or slower-declining unemployment rate in low-growth environment. While, the opposite is true in high-growth environment: women's unemployment rate declines faster or rises slower than men's, which is perplexed by their more intense passivation on the labor market.

**Figure 6** splits the within-country periods in monetary policy easing (negative monetary policy shock) and tightening (positive shock). Unexpected easing of monetary policy does not impinge on the gender employment gap, while the earlier observed worsening of the gap is likely mainly derived when interest rates have been unexpectedly raised. The same conclusion holds when youth are only considered, but the worsening of the gap comes sooner and then rebounds by the end of the period. However, even under monetary policy easing, women experience some worsening of their employment position around year two, which suggests that even if employment conditions under monetary-policy easing improve, then they do so more for men than for women. Labor force participation gap worsens under both easing and tightening, but the worsening under tightening is more endured until year four. Finally, gender unemployment gap worsens (to the advantage of women) under both easing and tightening, but the result under easing is not stable, while under tightening is fairly small.

Overall, as monetary policy shocks hits women in developing economies more strongly than men, then this hit is felt more intensively when the economy is growing faster than the global average, suggesting that monetary shock in a faster growing economy exacerbates gender inequality more so than in a slower growing economy. Likewise, unexpected monetary policy tightening hits women more severely than easing does; actually, the role of easing for employment is negligible if at all existing.

**Figure 5 – Gendered labor market response to a monetary policy shock, low- versus high-growth countries**

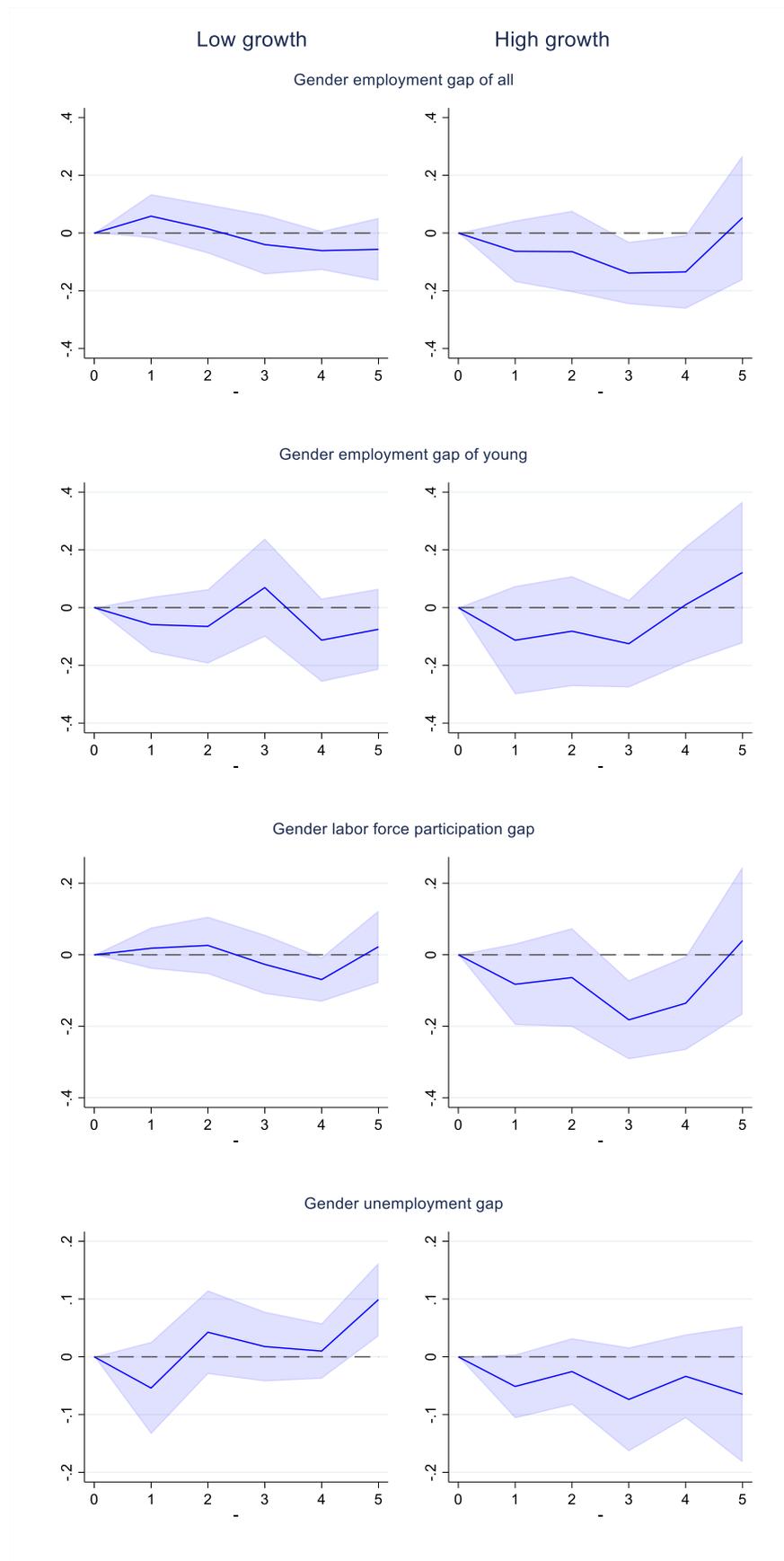

*Source: Authors' estimates.*

*Notes: Each graph presents the response of the titled labor market indicator to a one standard deviation monetary policy shock, separately for low growth countries (column 1), and high-growth countries (column 2). Hence, the vertical axis represents percentage points. The horizontal axis presents the time horizon expressed in years. 90% confidence interval is presented in shading. A positive (negative) impulse response represents a narrowing (widening) of the gender gap.*

# Figure 6 – Gendered labor market response to a monetary policy shock, negative versus positive monetary shocks

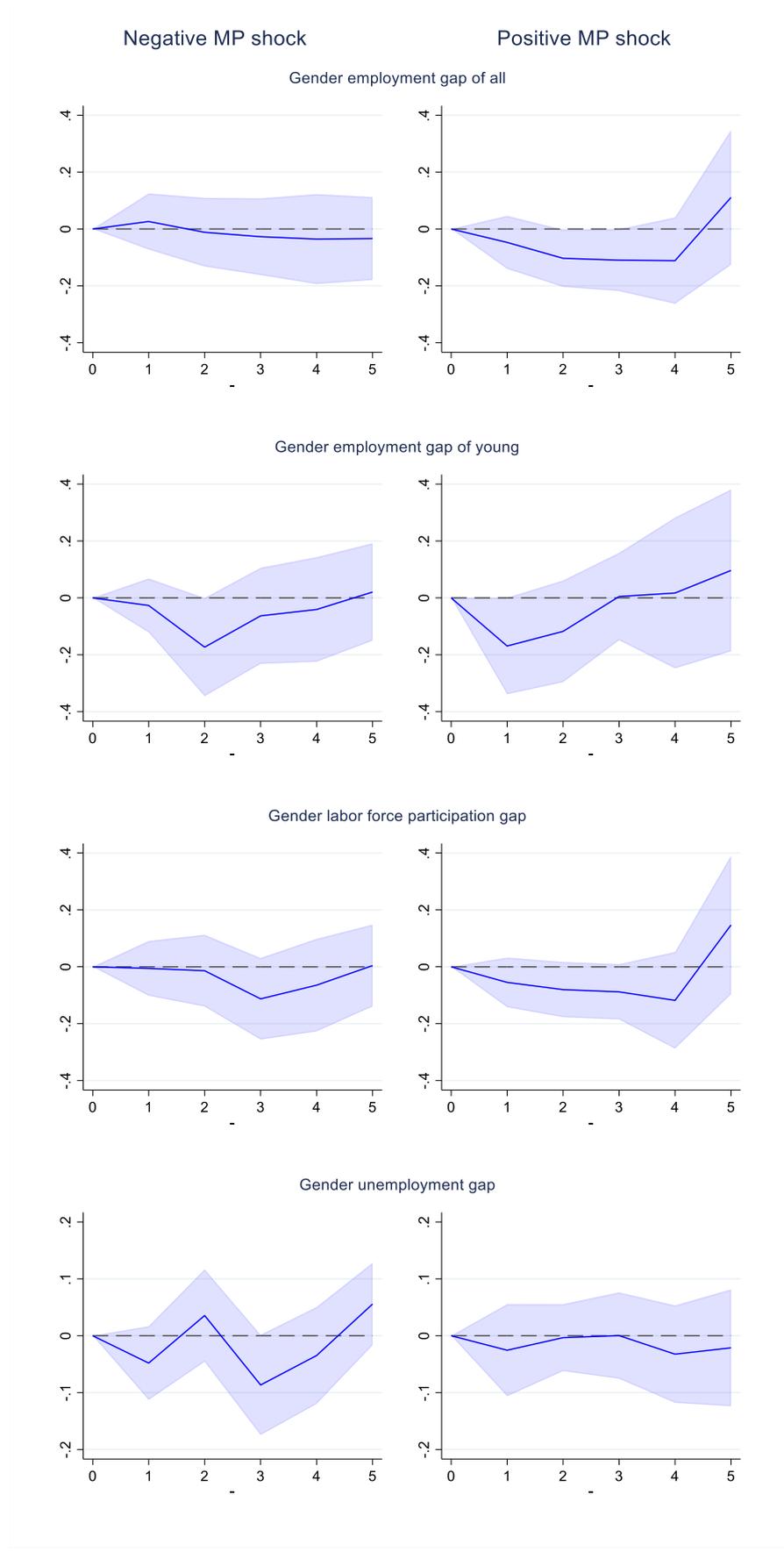

*Source: Authors' estimates.*

*Notes: Each graph presents the response of the titled labor market indicator to a one standard deviation monetary policy shock, separately for negative monetary policy shock (easing) (column 1), and positive monetary policy shock (tightening) (column 2). Hence, the vertical axis represents percentage points. The horizontal axis presents the time horizon expressed in years. 90% confidence interval is presented in shading. A positive (negative) impulse response represents a narrowing (widening) of the gender gap.*

### 4.5. Results with geographical heterogeneity

In the last analytical section, we disentangle the results by regions. We divide the developing world on four major regions: Emerging and Developing Asia (EDA), Emerging and Developing Europe (EDE), Latin America and the Caribbean (LAC), and Sub-Saharan Africa (SSA) and the results are presented in Figure 7. A monetary policy shock of one standard deviation is detrimental for women's employment more than for men's - hence worsening the gender employment gap in at least two developing regions: Asia and Latin America. There, the lowest point is achieved around years three or four following the shock, but in Asia the decline is recovered already in year five. In Europe, the shock leaves the gender employment gap largely unchanged, as well as in Africa, yet until year five.

Gender employment gap of youth likewise worsens in three of the four regions, excepting Latin America, in the first or second year following the shock, which is in line with the general observation. In Latin America, the worsening appears later, in year four, and with no recovery afterwards. On the contrary, the decline in the other three regions is recovered by the end of the horizon. Overall, in terms of employment outcomes, Latin America and Africa seem to lack any recovery of the worsening of women's position more than that of men.

Similar pattern is observed when the gender labor participation gap is observed. Women suffer more than men when a monetary policy shock hits, with more intense withdrawal from the labor market compared to men. The strongest effect occurs at about year three to five; it takes longest in Africa for the shock to materialize in worsened gender labor participation gap. However, the reaction in Europe is the mildest among the regions, i.e. reveals weakest passivation of women following the shock.

The gender unemployment gap turns out to be more negative, which is beneficial for women, at least in Asia, Latin America and Africa, in line with the baseline results. On the figure below, the deepening of the gender unemployment gap is of similar magnitude as in the overall result, it is just the scale is now larger, due to the graph of Europe, which reveals a very strong negative in year one and then a very strong positive reaction in year two. Overall, in Europe, the gender unemployment gap remains intact on average, which corroborates the weak-passivation effect of women there. In Africa, the gap likewise remains intact in the first four years after the shock, which reflects the notion that the shock neither affected labor force participation by year five. In this year, the unemployment situation for women becomes direr compared to men. Overall, the gender labor participation and unemployment gaps follow similar patterns across regions, with Europe exhibiting the mildest effects on women's labor force dynamics.

Figure 7 – Gendered labor market response to a monetary policy shock, by region

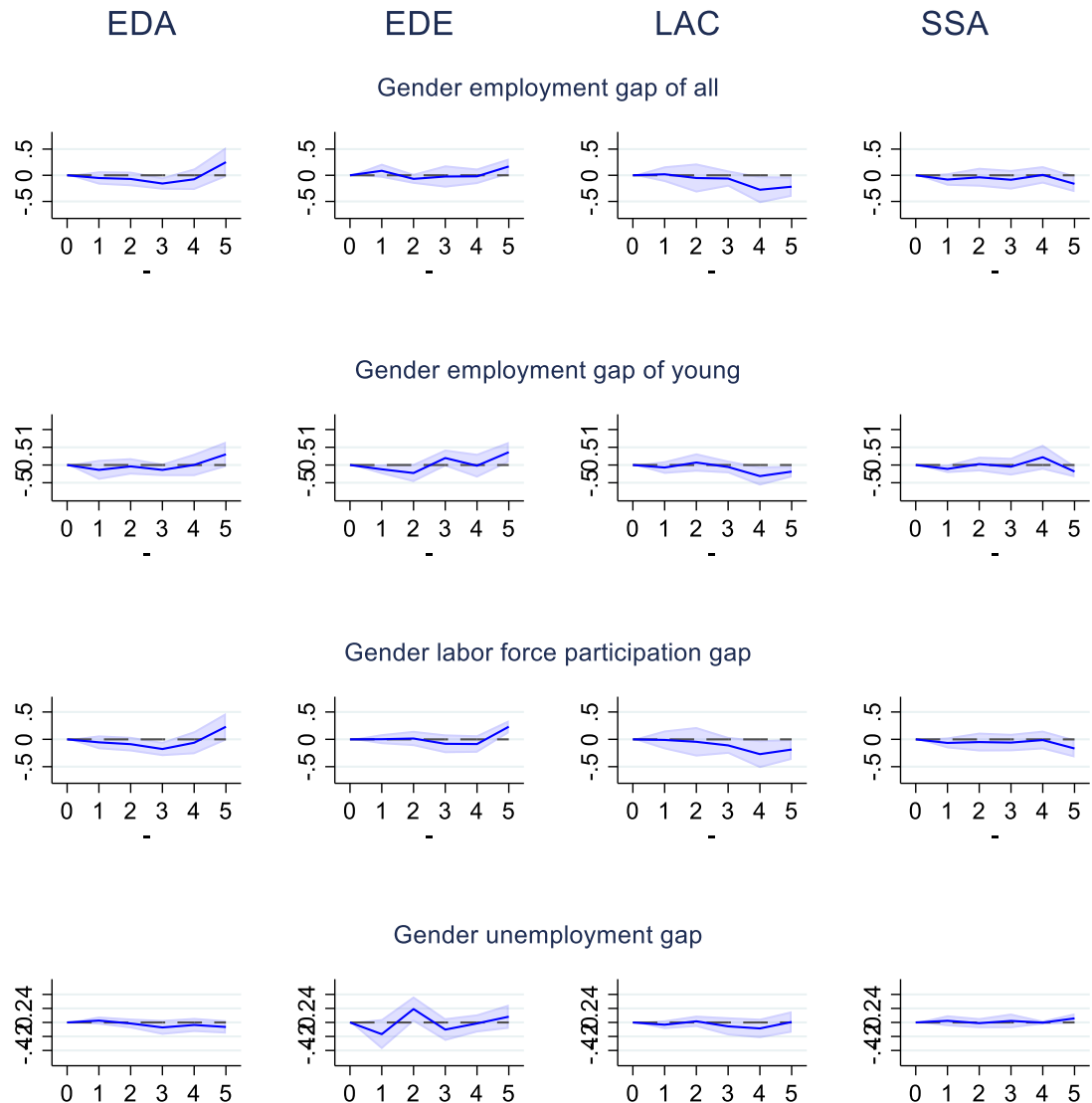

Source: Authors' estimates.

Notes: Each graph presents the response of the titled labor market indicator to a one standard deviation monetary policy shock. Hence, the vertical axis represents percentage points. The horizontal axis presents the time horizon expressed in years. 90% confidence interval is presented in shading. A positive (negative) impulse response represents a narrowing (widening) of the gender gap. Abbreviations stand for the world regions as follows: Emerging and Developing Asia (EDA), Emerging and Developing Europe (EDE), Latin America and the Caribbean (LAC), Sub-Saharan Africa (SSA).

## 5. Conclusion

The aim of this paper is to reveal if unanticipated monetary policy shocks may have implications for men's and women's labor-market outcomes in developing economies, by examining the specific impacts on gender gaps on the labor market in a set of 99 developing economies over the period 2009-2021.

To capture the rather inflexible forms in the exchange rate regimes of our sample, we firstly obtain the monetary policy shocks from a Taylor rule augmented with countries' official reserves movements. This tends to catch a common monetary policy reaction of the monetary authorities to usually avoid the instability of foreign exchange market. Secondly, these country-specific shocks are plugged into a local projections' estimator to understand how the gender-disaggregated labor-market outcomes react to a monetary policy shock of one standard deviation.

The baseline results reveal nuanced patterns. After a monetary policy shock, men generally experience a slightly positive response in employment outcomes, while women's employment reacts more negatively, particularly in the third and fourth years after the shock. This asymmetry contributes, in turn, to a gradual deepening of the gender employment gap, which peaks around the third or fourth year and diminishes thereafter. This outcome is complemented with deterioration of the gender participation gap, that is women more intensively passivize following a shock then men. Under such adjusting mechanism, the observed gender unemployment gap either remains unchanged or improves for women.

Including controls such as lagged GDP growth provides robustness to our findings. The results remain consistent, highlighting the persistent impact of monetary policy shocks on gender employment dynamics. Additionally, exploring alternative specifications of monetary policy shocks further corroborates the gendered outcomes, emphasizing the higher sensitivity of women's employment to unexpected policy changes.

The labor market adjustments post-policy changes exhibit sector-specific responsiveness to interest rates and are influenced by the structural characteristics of the labor market. Despite pronounced differences, particularly in men's outcomes, arising from labor market adjustment heterogeneity and data availability limitations, the results remain robust. Notably, sectoral reactions following a monetary policy shock vary: agriculture declines and industry and services improve. This nuances the initial assumption of sectoral impacts on women, suggesting that gender gaps persist despite sector-specific adjustments. The labor force participation and unemployment rate responses are more consistent, with women initially experiencing increased unemployment, subsequently stabilizing the gender unemployment gap.

The gender employment gap worsens more significantly in high-growth environments, suggesting that the intersection of monetary shocks and rapid economic expansion exacerbates gender inequalities. Moreover, the adverse effects are more pronounced under monetary policy tightening than easing, indicating that unexpected hikes in interest rates disproportionately affect women's employment. This may imply, for example, that in common cases for many smaller development economies which expose their fixed currency to speculation, may ultimately lead to harm women on the labor market more significantly than men.

Geographical heterogeneity analysis reveals distinct regional patterns. Developing Asia and Latin America experience the most significant detrimental effects on women's employment, with a recovery observed in Asia by the fifth year. Europe shows the mildest impact, aligning with the high activity rates of women in the labor market. Africa exhibits a slower manifestation of the shock's impact on the gender employment gap, with the situation for women worsening by the fifth year.

The comprehensive analysis underscores the need for targeted policy interventions to mitigate the gendered consequences of monetary policy shocks in developing economies, which extend beyond monetary and structural (labor-market) policies to fiscal and redistribution policies. Policymakers should be attentive to the differential impacts on men and women, crafting measures that promote gender equality and resilience in the face of economic shocks. Particularly, if central bank start to gradually include understanding of the differentiated impact of the policy move by gender in their analytical approaches, this may help to the calibration of other policies – e.g. the active labor market policies, vocational training, fostering part-time work and other flexible arrangements, unemployment benefits – in the way that best suits affected groups, particularly women. Our research not only advances the understanding of the intricate relationship between monetary policy shocks and gender employment gaps but also underscores the urgency of adopting inclusive policy frameworks to ensure equitable outcomes in diverse economic contexts.

This study is one of the largest-possible panel of developing and emerging economies, given data limitations. This, as in every other panel study, assumes aggregation of the results. Hence, tailoring policies to the specific circumstances of each country is essential. Consequently, additional research is needed to enhance the dynamics and determinants of the gendered impacts arising from monetary policy on the local labor markets, considering factors like the level of economic and social development, the very specifics of monetary policy design and execution, and the structures of the labor market.

## Appendix 1

The Appendix provides estimates of the Taylor rule (equation (1) in the main text of this paper). While what we need is the country-specific estimates, here we provide panel-based estimates to observe the Taylor rule in developing economies as a group. We provide results of a simple FE estimator (columns 1 and 2), IV estimates (columns 3 and 4) and Arellano-Bond estimates (columns 5 and 6), all in **Table A 1**. The idea with the latter two groups of estimates is to allow for capturing any remaining endogeneity in the model, despite we intentionally take the first lags of the non-forecast variables. However, some endogeneity may be still present, for example through undertaking some investment decisions given expectations about the future interest rates, especially when financing sources have pronounced component of the interest rate tied to the central-bank reference rate.

Results for the Taylor rule have the expected signs, except the forecasted GDP growth rate. Lagged GDP growth is positively related to the interest rate, reflecting the usually observed relationship. However, higher expected growth is predicted to result in lower interest rate, which is counter-intuitive. Still, both results on GDP-interest rate relationship are not stable across specifications. This is not the case for the inflation rate: both lagged and forecasted one robustly lead to increasing interest rate. While, a decline in reserves results in an increase in the nominal interest rate, reflecting an attempt to curb inflationary pressures or stabilize the currency. Finally, negative lagged interest rate, while strange at first, reflects the notion that higher previous levels of the interest rate lead to smaller subsequent changes in the nominal interest rate.

Table A 1 – Panel estimates of the Taylor-type of rule

|  | \multicolumn{6}{c}{*Dependent variable: Changes in the nominal interest rate*} | | | | | |
|---|---|---|---|---|---|---|
|  | FE estimates | | IV estimates | | Arellano-Bond estimates | |
|  | (1) | (2) | (3) | (4) | (5) | (6) |
| Lagged GDP growth | 0.0253** | 0.00552 | 0.186 | 0.211 | 0.0321** | 0.00408 |
|  | (0.011) | (0.017) | (0.144) | (0.266) | (0.015) | (0.022) |
| Forecasted GDP growth | -0.0344*** | -0.0261** | -0.326 | -0.439 | -0.037 | -0.0231** |
|  | (0.012) | (0.012) | (0.283) | (0.494) | (0.036) | (0.012) |
| Lagged inflation rate | 0.0204*** | 0.0205** | 0.0662** | 0.0594 | 0.0152** | 0.0133*** |
|  | (0.007) | (0.008) | (0.032) | (0.038) | (0.006) | (0.005) |
| Forecasted inflation rate | 0.125** | 0.127** | 0.424*** | 0.394*** | 0.0495* | 0.0396** |
|  | (0.062) | (0.064) | (0.130) | (0.144) | (0.027) | (0.020) |
| Lagged changes in reserves | -0.00730*** | -0.00556* | -0.0361 | -0.077 | -0.00668** | -0.0042 |
|  | (0.002) | (0.003) | (0.082) | (0.105) | (0.003) | (0.003) |
| Lagged nominal interest rate | -0.185* | -0.203* | -0.645*** | -0.602*** | -0.185*** | -0.274*** |
|  | (0.095) | (0.117) | (0.123) | (0.158) | (0.072) | (0.072) |
| Constant | 1.352 | 1.516 |  |  |  |  |
|  | (1.354) | (1.930) |  |  |  |  |
| Time dummies | No | Yes | No | Yes | No | Yes |
| Observations | 975 | 975 | 743 | 743 | 856 | 856 |
| Nb. of countries | 99 | 99 | 96 | 96 | 99 | 99 |
| Hansen test |  |  | 0.926 | 0.957 | 0.156 | 0.0923 |

*Source: Authors' calculations. *, ** and *** denote a statistical significance at the 10%, 5% and 1% level, respectively. Standard errors are robust to heteroskedasticity, are clustered, and provided in parentheses.*

## Appendix 2

### Table A 2 – Variables and their sources

| Variable | Description | Source |
|---|---|---|
| Interest rate | The bank rate that usually meets the short- and medium-term financing needs of the private sector. It is used both in its level (lagged value) and change compared to the previous period. | International Financial Statistics |
| GDP growth | GDP growth rate in real terms. It is used in its lagged value | World Economic Outlook |
| GDP growth forecast | One-year ahead forecast of the GDP growth rate in real terms | World Economic Outlook |
| Inflation rate | Average inflation rate. It is used in its lagged value | World Economic Outlook |
| Inflation forecast | One-year ahead forecast of the average inflation rate | World Economic Outlook |
| Change in reserves | Annual change in holdings of monetary gold, special drawing rights, reserves of IMF members held by the IMF, and holdings of foreign exchange under the control of monetary authorities. It is used in its lagged value | International Financial Statistics |
| Employment rate | Employment to population ratio, 15+, male (%) | International Labor Organization |
| | Employment to population ratio, 15+, female (%) | |
| | Employment to population ratio, ages 15-24, male (%) | |
| | Employment to population ratio, ages 15-24, female (%) | |
| Share of employment in agriculture | Employment in agriculture, female (% of female employment) | International Labor Organization |
| | Employment in agriculture, male (% of male employment) | |
| Share of employment in industry | Employment in industry, female (% of female employment) | International Labor Organization |
| | Employment in industry, male (% of male employment) | |
| Share of employment in services | Employment in services, female (% of female employment) | International Labor Organization |
| | Employment in services, male (% of male employment) | |
| Labor force participation rate | Labor force participation rate, male (% of male population ages 15+) | International Labor Organization |
| | Labor force participation rate, female (% of female population ages 15+) | |
| Unemployment rate | Unemployment, male (% of male labor force) | International Labor Organization |
| | Unemployment, female (% of female labor force) | |
| Collective bargaining | Collective bargaining coverage rate (%) | International Labor Organization |
| Gender pay gap | Average hourly earnings of employees of women (Local currency) minus average hourly earnings of employees of men (Local currency), divided by the former | International Labor Organization |
| Share of informal employment | Proportion of informal employment in total employment | International Labor Organization |

Table A 3 – Descriptive statistics of the included variables

| Variable | Obs. | Mean | St.dev. | Min. | Max. |
|---|---|---|---|---|---|
| Interest rate | 983 | 12.19 | 7.33 | 1.47 | 67.25 |
| Inflation rate | 983 | 5.67 | 14.91 | (3.09) | 379.85 |
| Forecast of the inflation rate | 983 | 5.32 | 6.41 | (1.00) | 110.69 |
| Real GDP growth rate | 983 | 3.03 | 5.04 | (33.50) | 43.48 |
| Forecast of the real GDP growth rate | 983 | 4.42 | 4.21 | (6.07) | 85.62 |
| Change in the interest rate | 983 | (0.32) | 1.92 | (12.69) | 21.94 |
| Change in reserves | 983 | 7.36 | 21.17 | (70.35) | 180.82 |
| Monetary policy shocks | 875 | 0.00 | 0.72 | (5.70) | 5.92 |
| Standardizes monetary policy shocks | 864 | (0.00) | 0.95 | (2.81) | 2.63 |
| Employment rate of men (15+) | 924 | 67.43 | 10.74 | 40.58 | 96.28 |
| Employment rate of women (15+) | 924 | 46.18 | 13.31 | 9.71 | 82.29 |
| Employment share in agriculture, women | 922 | 25.49 | 22.98 | 0.03 | 86.70 |
| Employment share in agriculture, men | 922 | 28.23 | 16.66 | 1.35 | 76.28 |
| Employment share in industry, women | 922 | 12.17 | 6.96 | 0.95 | 42.32 |
| Employment share in industry, men | 922 | 25.56 | 9.07 | 6.21 | 63.14 |
| Employment share in services, women | 922 | 62.34 | 21.89 | 11.64 | 96.11 |
| Employment share in services, men | 922 | 46.21 | 11.28 | 14.81 | 76.64 |
| Employment rate of men (15-24) | 924 | 41.35 | 14.20 | 12.51 | 82.12 |
| Employment rate of women (15-24) | 924 | 28.03 | 13.08 | 3.97 | 71.25 |
| Labor force participation of men | 924 | 72.48 | 8.90 | 43.57 | 96.38 |
| Labor force participation of women | 924 | 50.44 | 13.26 | 12.27 | 83.90 |
| Unemployment rate of men | 924 | 7.27 | 5.69 | 0.05 | 32.85 |
| Unemployment rate of women | 924 | 8.93 | 6.47 | 0.24 | 33.57 |
| Share of informal employment | 380 | 56.32 | 23.00 | 3.84 | 97.07 |
| Collective bargaining coverage rate | 223 | 23.51 | 23.29 | 0.40 | 98.50 |
| Gender pay gap | 334 | (4.97) | 15.74 | (49.65) | 147.38 |

Table A 4 – Countries included

| Country | Years observed | Region (geo) |
|---|---|---|
| Albania | 8 | EDE |
| Algeria | 12 | SSA |
| Angola | 10 | SSA |
| Antigua and Barbuda | 12 | LAC |
| Argentina | 6 | LAC |
| Armenia | 12 | EDE |
| Azerbaijan | 12 | EDE |

| Country | | |
|---|---|---|
| Bahrain | 6 | EDA |
| Bangladesh | 11 | EDA |
| Barbados | 12 | LAC |
| Belarus | 12 | EDE |
| Belize | 12 | LAC |
| Bhutan | 6 | EDA |
| Bolivia | 10 | LAC |
| Bosnia and Herzegovina | 12 | EDE |
| Botswana | 9 | SSA |
| Brazil | 12 | LAC |
| Brunei Darussalam | 12 | EDA |
| Bulgaria | 12 | EDE |
| Cabo Verde | 9 | SSA |
| Chile | 9 | LAC |
| China | 12 | EDA |
| Colombia | 10 | LAC |
| Comoros | 10 | SSA |
| Costa Rica | 12 | LAC |
| Croatia | 5 | EDE |
| Democratic Republic of Congo | 5 | SSA |
| Dominica | 9 | LAC |
| Dominican Republic | 12 | LAC |
| Egypt | 12 | SSA |
| Fiji | 10 | EDA |
| Georgia | 12 | EDE |
| Grenada | 9 | LAC |
| Guatemala | 12 | LAC |
| Guyana | 10 | LAC |
| Haiti | 12 | LAC |
| Honduras | 12 | LAC |
| Hungary | 12 | EDE |
| India | 13 | EDA |
| Indonesia | 12 | EDA |
| Jamaica | 12 | LAC |
| Jordan | 7 | SSA |
| Kenya | 12 | SSA |
| Kosovo | 7 | EDE |
| Kuwait | 11 | EDA |
| Kyrgyz Republic | 12 | EDE |
| Lesotho | 8 | SSA |
| Liberia | 5 | SSA |
| Madagascar | 6 | SSA |
| Malaysia | 12 | EDA |
| Maldives | 8 | EDA |
| Mauritius | 12 | EDA |
| Mexico | 12 | LAC |
| Moldova | 12 | EDE |
| Mongolia | 10 | EDA |
| Montenegro | 10 | EDE |
| Mozambique | 12 | SSA |
| Myanmar | 11 | EDA |
| Namibia | 8 | SSA |
| Nicaragua | 12 | LAC |
| Nigeria | 12 | SSA |
| North Macedonia | 12 | EDE |

| Country | | |
|---|---|---|
| Oman | 9 | SSA |
| Pakistan | 12 | EDA |
| Panama | 12 | LAC |
| Papua New Guinea | 5 | EDA |
| Paraguay | 12 | LAC |
| Peru | 11 | LAC |
| Philippines | 10 | EDA |
| Qatar | 12 | SSA |
| Romania | 12 | EDE |
| Rwanda | 12 | SSA |
| Samoa | 12 | SSA |
| Seychelles | 10 | SSA |
| Sierra Leone | 10 | SSA |
| Solomon Islands | 9 | SSA |
| South Africa | 10 | SSA |
| South Sudan | 5 | SSA |
| Sri Lanka | 10 | SSA |
| St. Kitts and Nevis | 12 | LAC |
| St. Lucia | 12 | LAC |
| St. Vincent and the Grenadines | 12 | LAC |
| Suriname | 8 | LAC |
| Tajikistan | 10 | SSA |
| Tanzania | 6 | SSA |
| Thailand | 12 | EDA |
| The Gambia | 7 | SSA |
| Timor-Leste | 8 | EDA |
| Tonga | 7 | SSA |
| Trinidad and Tobago | 9 | LAC |
| Uganda | 9 | SSA |
| Ukraine | 12 | EDE |
| Uruguay | 10 | LAC |
| Uzbekistan | 8 | EDA |
| Vanuatu | 5 | SSA |
| Vietnam | 12 | EDA |
| Zambia | 11 | SSA |

Note: Abbreviations stand for as follows: Emerging and Developing Asia (EDA), Emerging and Developing Europe (EDE), Latin America and the Caribbean (LAC), Sub-Saharan Africa (SSA).